# The Local Field Potential Reflects Surplus Spike Synchrony

Michael Denker[1], Sébastien Roux[2], Henrik Lindén[3], Markus Diesmann[1], Alexa Riehle[2], Sonja Grün[1,4]

[1] *RIKEN Brain Science Institute, Wako City, Japan*

[2] *Mediterranean Institute of Cognitive Neuroscience (INCM), CNRS - University Aix-Marseille 2, Marseille, France*

[3] *Dept. of Mathematical Sciences and Technology, Norwegian University of Life Sciences, Ås, Norway*

[4] *Bernstein Center for Computational Neuroscience, Berlin, Germany*

# Abstract

The oscillatory nature of the cortical local field potential (LFP) is commonly interpreted as a reflection of synchronized network activity, but its relationship to observed transient coincident firing of neurons on the millisecond time-scale remains unclear. Here we present experimental evidence to reconcile the notions of synchrony at the level of neuronal spiking and at the mesoscopic scale. We demonstrate that only in time intervals of excess spike synchrony, coincident spikes are better entrained to the LFP than predicted by the locking of the individual spikes. This effect is enhanced in periods of large LFP amplitudes. A quantitative model explains the LFP dynamics by the orchestrated spiking activity in neuronal groups that contribute the observed surplus synchrony. From the correlation analysis, we infer that neurons participate in different constellations but contribute only a fraction of their spikes to temporally precise spike configurations, suggesting a dual coding scheme of rate and synchrony. This finding provides direct evidence for the hypothesized relation that precise spike synchrony constitutes a major temporally and spatially organized component of the LFP. Revealing that transient spike synchronization correlates not only with behavior, but with a mesoscopic brain signal corroborates its relevance in cortical processing.

# Introduction

It is common belief that the local field potential (LFP), a population signal obtained from electrophysiological recordings of the brain, should reflect the synchronized spiking activity of neurons in the vicinity of the recording electrode. This assumption is rooted in the widely accepted biophysical explanation of the LFP as a spatially weighted average of the synaptic transmembrane currents (Mitzdorf, 1985; Viswanathan and Freeman, 2007). Indeed, the average postsynaptic effect in the LFP at a given recording site triggered on spikes initiated across a patch of cortex is predictive of the LFP (Nauhaus et al., 2009). In consequence, the oscillatory structure observed ubiquitously in the LFP is hypothesized to reflect predominantly oscillatory synchronized input (Logothetis and Wandell, 2004). Indeed, the LFP has been shown to correlate with membrane potential oscillations of nearby neurons (Poulet and Petersen, 2008) independent of the spiking activity (Okun et al., 2010). However, although the extension from membrane potential dynamics to coincident spiking activity is on everybody's mind, the hypothesis that synchronized action potentials are reflected in LFP oscillations has not been directly shown.

A large body of literature investigates the relationship of spikes and the LFP. To date, it has been established that neural spiking activity may become transiently coupled to the LFP in a rhythmic or non-oscillatory fashion (Eckhorn and Obermueller, 1993; Murthy and Fetz, 1996b). The degree of phase locking between neurons and the LFP depends in general on the strength of beta/gamma LFP oscillations (Denker et al., 2007), and both auto-correlations and cross-correlations between simultaneously recorded neurons tend to show an oscillatory structure during strong oscillatory episodes (Murthy and Fetz, 1996b). Such oscillatory periods are correlated with stimulus features (Engel et al., 1990) as well as top-down processes, such as attention (Fries et al., 2001), and are thus believed to be computationally informative (Fries et al., 2007). Indeed, firing rate profiles correlate with gamma band LFP power when the level of interneuronal rate correlation is high (Nir et al., 2007), and the power correlation between the spiking activity of different neuronal groups depends crucially on their phase relationship with the LFP (Wommelsdorf, 2007). In addition, a number of studies indicate that across brain areas, inhibitory neurons play a crucial role in the generation of fast oscillations (Klausberger et al., 2003; Hasenstaub et al., 2005; Cardin, 2009). Excitatory-inhibitory loops (Berens et al., 2008) gate the temporal structure of activity projecting onto pyramidal cells (Buzsáki and Draguhn, 2004).

Despite the fact that oscillatory activity in the LFP is reflected on the level of membrane potentials and rate co-modulations, it remains unclear how the LFP oscillation is related to the precise synchronization of individual action potentials. Recent studies succeeded to directly relate synchronized slow subthreshold membrane potential oscillations to LFPs, but did not find such a relationship for synchronized action potentials of the same neurons (Poulet and Petersen, 2008). This discrepancy between subthreshold dynamics and spiking activity is in agreement with theoretical work linking subthreshold and suprathreshold dynamics (Tetzlaff et al., 2008). In consequence, the findings of Poulet and Petersen (2008) indicate that the occurrence of action potentials is governed by strong, precisely timed, and specific inputs to the cells suggesting these as independent activity riding on the co-modulating oscillations. Moreover, a recent study by Okun et al. (2010) questions the idea that network-wide population events dominate the LFP, suggesting that precise firing occurs in smaller groups of neurons, and therefore might only be subtly represented in the LFP.

.

One hypothesis that is compatible with such input characteristics states that specific common inputs force the precise synchronous discharge within a defined group of cells, termed the Hebbian cell assembly (Hebb, 1949). Early on, it has been conjectured that LFP oscillations may represent an alternative network-averaged signature of assembly activations (Donoghue et al., 1998; Singer, 1999) and enable the binding of features coded by different assemblies (Eckhorn et al., 1988). Indeed, distinct spike patterns across neurons and their phase relationship to LFP oscillations encode a substantial amount of surplus of information about the stimulus compared to information contained in the firing rate alone (Kayser et al., 2009). Nevertheless, the critical link between the dynamics of precise interneuronal spike correlations and the LFP on a trial-by-trial basis is missing. In particular in motor cortex, there is no intuitive correspondence between spatially extended (Fig. 1A; cf. also Rubino et al., 2006) LFP oscillations and spike synchronization in the absence of a network oscillation in the spiking activity (Fig. 1B-E; cf. also Nawrot et al., 2008).

On the spiking level, the hallmark signature of an activated assembly is the functionally coordinated synchronous spiking with millisecond precision observed in parallel recordings of neuronal activity (Gerstein et al., 1989) that exceeds the expectation based on the neuronal firing rates (Aertsen et al., 1989). It is shown that not only LFP oscillations correlate with external stimuli (e.g., Montemurro et al., 2008), behavioral aspects (e.g., Scherberger et al., 2005), and internal processes (e.g., Murthy and Fetz, 1996a; Donoghue et al., 1998; Roux et al., 2006), but also precise spike synchrony is observed (Riehle et al., 1997; Vaadia et al., 1995) and modulated (Kilavik et al., 2009) in a functional context. For beta/gamma oscillations it remains an open question if LFPs reflect more than synchronization due to an underlying rate modulation, and if these oscillations may provide a framework for the occurrence of precisely coordinated spiking as predicted by an active assembly (Buzsáki, 2004; Jensen, 2006). Here, we uncover this missing link between observed spike synchrony and LFP oscillations by directly relating these observables.

## Materials and Methods

### Ethics Statement

Care and treatment of the animals during all stages of the experiments conformed to the European and French government regulations, according to the Weatherall report ('The use of non-human primates in research', December 2006).

### Experimental design and electrophysiological recordings

All data were taken from recordings partially presented elsewhere (Roux et al., 2006; Kilavik et al., 2009). Two rhesus monkeys (monkey K and monkey O) were trained to perform arm movements from a center position to one of two possible peripheral targets left and right of the center in two different tasks involving an instructed delay. In the first, a choice reaction time task (chRT), both peripheral targets were presented simultaneously as a preparatory signal (PS), one in red and the other in green. The animal learned to attribute to each color one of two possible delay durations. If the (directionally non-informative) auditory response signal (RS) occurred after a short delay, the monkey had to select the red target, after a long delay the green one. Both the laterality of the colored targets and the presentation of the two durations were varied at random with equal probability. In contrast, in the second self-paced movement task (SELF), the presentation of only one peripheral target, either in red or green, either at the left or the right, required a self-initiated response after estimating one of the two delays as coded by PS. In both tasks (Roux et al., 2006), four different timing patterns were used to identify the short and long delay, respectively: (i) 500 ms and 1000 ms (monkey K); (ii) 500 ms and 1200 ms (monkey K); (iii) 600 ms and 1200 ms (monkey O); (iv) 1000 ms and 1400 ms (monkey O).

In this study we exclusively analyzed the delay activity, i.e. activity recorded during the preparatory period (PP) starting at PS and ending with either RS in the chRT task or the earliest allowed response time (AT) in the SELF task. Therefore, the trials were aligned to PS occurrence for the analysis. The neural activity related to movement execution, i.e. after RS or AT, respectively, is not analyzed. For both tasks, only correct trials were considered, in which the monkey responded within a time window (after the end of PP) of maximally 300 ms (monkey O) and 500 ms (monkey K) and in which movements were performed in the required movement direction.

In order to exclude effects due to pooling of neuronal activities of different behavioral contexts and different tasks, their activity was analyzed separately for the four possible behavioral conditions (combinations of short or long delay duration and left or right upcoming movement direction) and each experimental session. For the sake of simplicity, we refer in this manuscript to a recorded neuron by the combination of its identity and the behavioral context during which it was recorded. In this sense, data recorded from the same neuron may enter a population average up to eight times (maximum of four different conditions in two tasks).

## Data acquisition and data analysis

LFPs and spikes were recorded simultaneously in primary motor cortex using a multielectrode device of 2-4 electrodes (MT-EPS, Alpha Omega). Spikes of single neurons were detected by an online sorting algorithm (MSD, Alpha Omega, Nazareth, Israel). The inter-electrode distance was on the order of 400 μm. LFPs were sampled at a resolution of 250-500 Hz and hardware filtered (band pass, 1-100 Hz). In total, we analyzed 53 recording sessions (monkey K: 25; O: 28), which yielded 143 single neurons or 570 combinations of neurons and behavioral conditions. On average 33±11 trials were recorded per experimental condition. In analyses that combine spikes and LFP, each neuron enters only once, and we never combined LFP and spikes that were recorded on the same electrode to exclude the possibility of spike artifacts in the signal. We confirmed that simultaneously recorded LFPs are highly synchronous in the frequency regimes of interest. Likewise, coincident activity between neurons was analyzed only from neurons recorded from different electrodes, totaling 123 analyzed pairs of neurons. All data analysis was performed using the Matlab software environment (The Mathworks Inc., Nattick MA).

## Coincidence detection and Unitary Event Analysis

From simultaneously recorded spike data of individual sessions we extract all unique pair combinations of spike trains that are recorded from distinct electrodes. In a first step, we compute the number of coincident spike occurrences of the pairs of neurons in a time-dependent manner (compare supplemental Fig. S1). To allow coincidences with a temporal jitter up to a maximal coincidence width of $b$=3 ms, we apply the 'multiple-shift' approach (Grün et al., 1999; Grammont and Riehle, 2003). In this method exact coincidences (within the time resolution $h$=0.1 ms of the data) are detected for a range of shifts between $-b$ to $+b$ of the second spike train against the first (reference) spike train. To account for the non-stationarity of the neurons' firing rates, and to capture the dynamics of correlation, we perform the Unitary Event (UE) analysis in a sliding window fashion (Grün et al., 2002b). This is done by moving a window of fixed duration (here: $T_w$=100 ms) along the data to cover the duration of a trial, i.e. the duration of the PP. The length of the time window is chosen large enough to include at least one complete cycle of the beta oscillation. The window is advanced in steps corresponding to the time resolution $h$ of the data. The first window position is centered at trial onset, and the last window at the end of the delay period.

Within each window position the total number of empirical coincidence counts $n_{\text{emp}}$ is derived by summing the exact coincident spike events from each shift $l$ and from all $M$ trials $j$: $n_{\text{emp}} = \sum_{j=1}^{M} \sum_{l=1}^{L} n_{\text{emp}}^{j,l}$, with $L=2(b/h)+1$. To derive UEs this count is compared to the number of coincidences that would occur by chance given the firing rates of the neurons. This involves the following calculations. To account for non-stationary rates across trials (Grün et al., 2003), the relevant measures are obtained from the single trial and only subsequently summed across trials. Thus, within the analysis window the expected number of coincidences is calculated on the basis of the trial by trial firing probabilities $p_{i,j}$ which are estimated by the spike count $c_{i,j}$ of neuron $i$ in trial $j$ divided by the number of bins $N$ within a window: $p_{i,j}=c_{i,j}/N$ with $N=T_w/h$. The joint probability for finding a coincidence by chance per trial is calculated by the product of the single neuron firing probabilities $p_{12,j}=p_{1,j}p_{2,j}$. The expected number of coincidences per trial $j$ results from multiplying this probability with the number of bins $N$ that are included in the analysis window and the number of shifts $L$: $n_{\exp}^{j} = NL\, p_{12,j}$. The total number of expected coincidences within the window is derived from the sum of the expected numbers per trial: $n_{\exp} = \sum_{j=1}^{M} n_{\exp}^{j}$.

Finally we compare the empirical $n_{\text{emp}}$ to the expected number $n_{\exp}$ of coincidences to detect significant deviations. To this end, we calculate the joint-p-value $jp$, i.e. the probability of measuring the given number of empirical coincidences (or an even larger number) under the null-hypothesis of independent firing. The distribution under this null-hypothesis representing the probability to find a given number of coincidences is given analytically assuming Poisson processes (Grün et al., 1999). The latter assumption is shown to yield a conservative estimate for cortical spike trains considering their non-Poisson and non-renewal properties (Grün, 2009). Then the significance of $n_{\text{emp}}$ yields (Grün et al., 2002a): $jp(n_{\text{emp}} | n_{\exp}) = \sum \frac{n_{\exp}^{r}}{r!} e^{-n_{\exp}}$. If its value is below an *a priori* threshold (here chosen as 5%) coincident firing is

classified as significant and identified as Unitary Events. Spikes are labeled as UE if they are part of at least one sliding window identified to contain significant excess synchrony (for an illustrated summary of this analysis approach, see Maldonado et al., 2008). In addition, we require such time windows to exhibit a minimum firing rate of 5 Hz for each neuron. Spikes that are part of coincident events but not identified as UE with respect to any of the neurons recorded in parallel are labeled as chance coincidences (CC), all remaining spikes as isolated spikes (ISO).

## Spectral analysis

Power spectra are used to assess the dominant frequencies in the LFP during the task. All power spectra are calculated using a Hamming window as taper. To illustrate the temporal modulation of power in different frequency bands, we use a time-resolved spectral analysis using 200 ms windows with a 50 ms overlap.

## Spike-triggered averages

Spike-triggered averages (STAs) are computed by averaging LFP segments from time windows of 200 ms centered at each spike time. For the STA analysis, LFPs are filtered between 2-80 Hz to remove DC components. To compare STAs across recordings, in which electrode signals often differ in their absolute amplitude values, we z-transform each LFP before further analysis by subtracting its mean (calculated across trials) and dividing by its standard deviation. In order to quantify the magnitude (or size) of an STA, we calculate the total area the STA encloses with the time axis. Similar results to those presented here (not shown) are obtained using alternative measures of the STA magnitude, such as the area under its envelope, or the maximum of its absolute value. The magnitude of the STA is in general dependent on the number of trigger spikes. In order to compare STAs obtained from two sets of trigger spikes of different number of spikes $n_1$ and $n_2$ ($n_1$>$n_2$) we construct 1000 STAs of set 1, each computed from $n_2$ randomly selected spikes. We define the STA of set 2 to be larger than that of set 1 if the magnitude of set 2 exceeds 50% of the re-computations of set 1, and significantly larger (at a level of 5%) if it exceeds 95% of the re-computations.

## Peak-triggered spike histograms

We evaluate the population-averaged spiking discharge triggered on the peaks of the LFP oscillation (Destexhe et al., 1999). To this end we detect maxima of the LFP separated by a minimum time interval of 33 ms, which corresponds to a maximal oscillation frequency of 30 Hz. The spike histogram is calculated from data within a window of 200 ms around each peak, and averaged across all individual peaks in all neurons (see Eeckman and Freeman, 1990 for a different technique to relate spike times to EEG time course based on amplitude). Simultaneously, we also compute the peak-triggered LFP by averaging the z-transformed LFP aligned on its peaks.

## Rate-amplitude correlation

To assess the degree of correlation between LFP oscillation strength and spike rates,

we calculate the mean value of the rectified, z-transformed LFP along each trial with sliding windows of 200 ms length and 100 ms overlap. These values are then correlated with the rate profile of the neuron estimated as the spike count across trials in the same windows. Similar results as those shown here are obtained using alternative measures of LFP strength, including the mean value of the envelope of the beta-filtered signal (compare phase-locking analysis), or by using the total signal power in the beta range (10-22 Hz).

**Phase analysis**

After examination of the dominant beta frequencies on a session-by-session basis, LFPs of both monkeys are filtered with a zero-phase 10-22 Hz band pass filter (Butterworth, 8-pole). Short filter transients in the time domain allow for good estimates of the instantaneous LFP amplitude. In a subsequent step, we calculate the instantaneous phase of the LFP from the analytic signal $\xi(t) = x(t) + i\,\tilde{x}(t)$ obtained via the Hilbert transformation $\tilde{x}(t) = \frac{1}{\pi}\,\mathrm{P.V.}\int \frac{x(t)}{t-\tau}\,d\tau$ of the original signal $x(t)$, where P.V. denotes that the integral is to be taken as Cauchy principal value (Le Van Quyen et al., 2001). In this formalism, troughs of the LFP are identified by a phase of $\pi$. The calculation of the analytic signal can be applied to arbitrary signals, but its interpretation as instantaneous phase is difficult where either the signal amplitude becomes too small to discriminate the oscillation from background noise, or where the regular oscillation is disrupted (Boashash, 1992). To account for these effects, we discard phase values which violate the monotonicity of the phase time series or exhibit instantaneous phase jumps. To further corroborate our results, we exclude from our analysis those 10% of spikes per neuron that occur at the lowest LFP amplitudes.

We analyze the distributions of extracted phase values at the times of spike occurrences (Denker et al., 2007) using tools from circular statistics (Mardia and Jupp, 2000). The mean phase $\phi$ is obtained via the circular average $R\, e^{i\phi} = N^{-1} \sum e^{i\phi(t_i)}$, where $\phi(t_i)$ indicates the phase of the field potential at time $t_i$ of spike $i$. Furthermore, we utilize the transformation of the vector strength $R$ to the circular standard deviation $\sigma = \sqrt{-2 \log R}$ as a measure of the concentration of the phase distribution. For small values, $\sigma$ relates to the standard deviation of a normal distribution, whereas for flat distributions it behaves as $\sigma \to \infty$. In all phase analysis, we discard neurons that fire in total (across trials) less than 25 spikes.

Additionally, we employ two measures to quantify whether spikes recorded from individual neurons show a significant phase preference to the LFP. For the first, we test against the null hypothesis that the phase sample is taken from the uniform circular distribution (Rayleigh test, cf. Mardia and Jupp, 2000), which is expected by assuming a regular (e.g., filtered) field potential and independent random spiking. However, spike trains that have a certain regular structure in time may display intrinsic locking to the LFP. To measure the degree of genuine locking that is not explained by the regularities of the two signals, we calculate as the second measure the degree of locking $R$ in 1000 surrogates, each created by shuffling the inter-spike intervals of the spikes on a trial-by-trial basis (random placement of the first spike).

This procedure preserves to first order the regularity manifested in the inter-spike interval distribution. A comparison with the measured value $R$ yields the p-value for this surrogate test. Since the construction of such surrogates can only be performed on the complete spike train, this measure could not be sensibly applied to the subsets of spikes in our analysis (i.e., ISO, CC, UE, as well as Lo and Hi used in the amplitude analysis).

The phase distribution of spike coincidences may be trivially sharpened due to a preferred phase occurrence of individual spikes. To correct for this effect we calculated the expected phase distribution of coincident spikes (compare black curve in Figs. 5 and 6). To this end, we calculate the joint phase probability distribution of a neuron pair by the phase-by-phase multiplication of the occurrence probabilities of spikes at these phases. The predictor for the whole population is the average of the pair-wise phase distributions weighted by the relative number of coincidences between the two neurons.

In contrast to this predictor which considers the phase of spikes irrespective of the spike interval distribution, we also construct a predictor based on the reverse scenario. For each pair of simultaneously recorded neurons the inter-spike intervals of the spike trains of each neuron are shuffled on a trial-by-trial basis to create a set of 1000 surrogate pairs. For each surrogate, the variance $\sigma$ is evaluated separately for the resulting sets of non-coincident and coincident spikes. Thus, we obtain for each neuron the variances $\sigma$ of phase locking of coincident and non-coincident spikes for the original data and for the 1000 surrogates, allowing us to compare their distributions (Fig. 4).

# Results

**Synchrony based spike classification**

We analyze spike data of 143 single units and simultaneously recorded LFP data from motor cortical areas in two monkeys during the instructed delay (preparatory period, PP) of two motor tasks (see Methods). Both spike synchrony (Kilavik et al., 2009) and LFP oscillations in the beta band (Murthy and Fetz, 1996a) have been shown to be behaviorally relevant to movement preparation. LFPs and spikes were recorded from different electrodes spaced at 400 μm (for a schematic illustration, see Fig. 2) to exclude trivial signal correlations induced by volume conductance effects (cf., e.g., Katzner et al., 2009). Using the Unitary Events analysis (Grün et al., 2002a,b), we identify transient periods where the spiking activity of simultaneously recorded sets of neurons shows a surplus of coincidence events compared to the number expected on the basis of the firing rates. During these periods we attribute the excess synchrony to the synchronous firing of both observed neurons as part of a network process that activates a specific subset of neurons: the assembly (Fig. 2 depicts the spikes of two different assemblies in green and blue). Based on this detection of precise spike synchrony (Grün et al., 1999) between all neuron pairs of a given neuron we classify the spikes recorded from each neuron (**all spikes**) exclusively into one of three sets: **isolated spikes (ISO)**, **chance coincidences (CC)**, and **Unitary Events (UE)**. Spikes involved in pairwise coincidences (within 3 ms) are classified as CC if they occur during time periods where the observed coincidence rate is explained by the instantaneous trial-by-trial rates of the two involved neurons, and as UE if their number significantly exceeds the expectation (see Methods). In a given UE period a distinction between coincidences stemming from the activation of the assumed assembly and those due to chance is not possible. Therefore, a substantial fraction (see Discussion for an estimate) of coincidences in the UE group may be due to chance coincident spiking (e.g., the rightmost UE coincidence in Fig. 2). Spikes not classified as CC or UE with respect to any of the simultaneously recorded neurons (2-5) are classified as ISO. Consequently each spike is labeled according to the type of event it belongs to, and an individual spike train may contain spikes of different categories (compare gray, cyan, and red boxes in Fig. 2, respectively).

## The magnitude of spike-triggered LFP averages increases with synchrony

As a first step, Fig. 3*A* compares the spike-triggered averages (STAs) of the LFP for the three sets, where each STA is pooled across all neuron-LFP pairs. We observe that the magnitude of the STAs of both chance coincidences (left, cyan) and Unitary Events (middle, red) significantly exceed that of the isolated spikes (gray). Moreover, the spike-triggered average of UE is larger than that of CC (right). The oscillatory structure of the STAs exhibits a strong beta frequency component, and the STAs are typically centered on the downward slope of the oscillation cycle. Non-averaged, single-neuron STAs also exhibit these differences, but to a lesser degree (see supplemental Fig. S2*A* for a typical example). The reason for this is two-fold: First, individual pairs have a substantially higher sampling variance, especially considering the typically low number of UE spikes. Second, STA shapes result from the combination of three effects: instantaneous LFP frequency, spike-LFP phase locking and oscillation amplitude. Nevertheless, the STA increase, in particular for UE spikes, is observed in a significant number of single neurons of both monkeys (Fig. 3*B*) and is consistently more pronounced for experiments where we were able to evaluate a larger number of partner neurons $N_p$ for potential coincidences (Fig. 3*C*), thus better separating the CC and UE groups.

Two mechanisms could underlie the differences in the STAs: changes in LFP amplitude or changes in the locking between LFP and spikes. However, the LFP amplitude does not co-vary with spike rate (Fig. 3*D*). Therefore increased amplitudes and the disproportionate increase of the chance coincidence count during periods of elevated rates is an improbable cause of the STA increase for CC. In addition, spike histograms triggered on the peaks of the LFP oscillations (supplemental Fig. S2*B*) reveal that spikes do not only tend to prefer the falling phase, but also avoid the rising phase of the LFP. This suggests that the three sets of spikes differ in the degree of phase coupling to the LFP rather than in the accompanying amplitude of the LFP.

**Increased spike synchrony improves spike-LFP phase coupling**

Nevertheless, in order to clearly differentiate between these mechanisms, it is necessary to formally disentangle the dependence of spike timing on the amplitude of the LFP from its dependence on the phase. Fig. 4*A* explains the procedure (for details see Methods). For both monkeys we consistently observe a prominent beta oscillation (in both monkeys around 15 Hz) of the LFP during the preparatory period that stops with movement onset (Mvt). Therefore we focus on the beta frequency band and extract the instantaneous phase and amplitude (envelope) of the field potential for each spike time. Compared to the STA analysis, even individual neurons exhibit clear and specific differences between ISO, CC, and UE in both measures (Fig. 5, same example neuron as in Figs. 1 and 4). We are now prepared to study the two contributions in detail across the population.

Fig. 4*B* shows that across the population of neurons CC are systematically better locked (decreased circular standard deviation $\sigma$ of the phase distribution) than ISO, and UE better than CC. As a suitable reference value to compare the fraction of locked neurons in the 3 sets we extracted the average locking strength $\sigma_l$=1.98 obtained for those neurons that are significantly locked if *all* spikes are considered (surrogate test). In the following we investigate how the systematic differences in locking strength between the three sets of spikes are affected by the intrinsic spike-LFP relationship of the neurons, i.e. if a neuron in general tends to lock well to the LFP or not. Differentiating groups of strongly (39%) and weakly (61%) locked neurons (i.e., significantly locked and unlocked neurons considering all their spikes) does not introduce a bias by affecting the percentage of neurons that exhibit CC and UE (supplemental Fig. S3*A*). Both groups exhibit the same general pattern of locking in the three groups (supplemental Fig. S3*B*) shown in Fig. 4*B*. As expected, the percentage of neurons better locked than $\sigma_l$ in the ISO group differs considerably (53% vs. 6%, gray bars in supplemental Fig. S3*B*) between strongly and weakly

locked neurons. However, this difference between strongly and weakly locked neurons is less pronounced for CCs (63% vs. 32%) and further decreases for UEs (65% vs. 46%). The conservation of the locking of UE spikes in strongly and weakly locked neurons compared to the declines for ISO and CC hints at different dynamical origins of the spikes in CC and UE.

Fig. 4*C* confirms that individual neurons are consistent with the findings for population ratios (Fig. 4*B*). The scatter plots of the circular standard deviation reveal that in 71% of the recorded neurons CC spikes are better locked than ISO spikes, and in 85% of the neurons UE spikes are better locked than ISO spikes. Finally, in 68% of all neurons UE spikes are better locked to the LFP than CC spikes. In contrast to the experimental data, only 58% of surrogate spike trains that retain the original inter-spike interval statistics show an increase in phase locking for coincident spikes (outlined ellipse).

Because of the consistency in the population, in the following we focus on the phase locking of strongly locked neurons. The rationale is to reduce the differences in locking between the three sets of spikes to obtain a conservative estimate of the locking (supplemental Fig. S3*B*). Comparable results are obtained for the complete set of recorded neurons. The phase distributions in the top panels of Fig. 6*A* show that locking of spikes to the LFP is strongest for Unitary Events, and weakest for isolated spikes.

The phase distribution exhibited already by isolated spikes modulates the spiking probability in time. Given the high level of synchrony between LFPs (Fig. 1*A*), one may therefore argue that the increased modulation of the phase distribution of CC trivially results from the individual phase locking distributions of the two neurons forming the coincidence (predictor assuming independence of neurons, see Methods). Interestingly, the phase distribution of CC is indeed largely in agreement with this predictor (black curve in Fig. 6*A*), while that of UE is not. Hence, despite the impossibility to remove the substantial fraction of chance coincidences from the UE group, the locking of UE cannot be explained on the basis of the intrinsic phase locking of the neurons forming the coincidences.

## Magnitude of global oscillations influences spike locking

Earlier studies (Murthy and Fetz, 1996b; Denker et al., 2007) demonstrate that spikes occurring during periods of high LFP amplitudes exhibit a stronger locking to the LFP. At a given time the amplitude of the LFP oscillation is defined by its envelope (blue curves in Fig. 4*A*). To examine the dependence of spike locking on the amplitude of the LFP (Denker et al., 2007), we form two exclusive sets of spikes, termed 'Hi' and 'Lo', based on whether a spike occurs at an amplitude above or below a certain value, respectively (Fig. 7*A*). We account for the session-by-session variability of the LFP amplitude by defining the threshold $\theta$ in terms of the fraction of spikes an individual neuron contributes to the Lo category (Fig. 7*B*).

For threshold ranges between 0.2 and 0.8 we observe that the percentage of significantly locked neurons (Rayleigh test, $\alpha=0.05$) of the Hi set is only decaying slightly from 41% to 36% (Fig. 7*C*). This percentage is in the same range as the percentage of locked neurons considering all spikes (Fig. 4*B*). We emphasize that even for high thresholds, where only few spikes are included, the locking of neurons can be explained using Hi spikes only. In contrast, when considering spikes of the Lo set, the percentage of locked neurons starts at 9% and increases approximately linearly with $\theta$ at a much steeper slope, meaning that at increasingly higher amplitudes more and more spikes are included in the Lo set. This shows that locking of spikes to the local field potential is largely due to spikes that occur at high LFP amplitudes.

## Combined effects of synchrony and LFP amplitude

Combination of the previous results raises the question of whether coincidences, and in particular Unitary Events, predominantly occur at high LFP amplitudes. Fig. 6*A* (density plots) shows the number of spikes as a function of both LFP phase and amplitude for each of the three sets ISO, CC, and UE. Here, CC and UE occur at similar amplitudes as ISO, even though the amplitude distributions (left) reveal a small shift towards high amplitudes for CC and UE. The phase distributions (top panels), however, clearly show a progressive increase in the degree of phase locking from ISO to CC to UE. Finally, observing that UEs exhibit similar amplitudes as CC, we can ask the reverse question of whether at high amplitudes ISO, CC and UE still exhibit the systematic increase in locking. Fig. 6*B* shows that for the 50% of the spikes occurring at the largest LFP amplitudes (above black dashed line in Fig. 6*A*) the effect of improved phase locking for the UE group is strongly amplified. In contrast, the ISO and CC phase distributions do not change. This finding reveals that those coincidences in UE periods that are responsible for the increased locking of UE are those that occur during strong LFP oscillations.

## Discussion

In this report we explicitly reveal how the local field potential relates to precise excess spike synchrony in motor cortex. Spikes which are emitted at the same time as spikes of other neurons exhibit a better phase locking to the dominant beta-range LFP oscillation than those which occur in isolation. However, in time periods where the number of spike coincidences is at chance level, the quality of the locking is explained by a predictor assuming independence of the spikes constituting a coincidence. In contrast, the pronounced locking to the LFP in time periods with a significant excess of coincident spikes (Unitary Events) cannot be explained in this way. The probability of the occurrence of coincident spikes is only weakly coupled to changes in the magnitude of the LFP signal. Nonetheless, spikes that coincide with episodes of high LFP amplitudes are on average better locked to the LFP than those at low amplitudes. A separate analysis of these two factors, identified spike synchrony and LFP magnitude, demonstrates that both affect the strength of the spike-LFP coupling largely independent of each other. What conclusions about network dynamics and possible coding mechanisms do these results imply, in particular in the light of the distinctive role of Unitary Events?

Features of the LFP signal correlate with external stimuli (O'Leary and Hatsopoulos,

2006), behavioral aspects (Scherberger et al., 2005), internal processes (Murthy and Fetz, 1996a; Poulet and Petersen, 2008; Roux et al., 2006), and attentional modulation (Fries et al., 2001). In particular, several authors have elucidated the functional role of LFP oscillations in motor cortex in the beta and lower gamma range. These oscillations are only loosely correlated across trials, i.e. their phase is not time-locked to any external (e.g. stimulus) or internal (e.g. movement onset) event. Oscillatory beta range LFP activity in motor cortex is a unique feature of experimental protocols including a waiting period before movement execution and has been described in relation to attentional processes, movement preparation and motor maintenance (Donoghue et al., 1998; O'Leary and Hatsopoulos, 2006; Murthy and Fetz, 1992, 1996a; Baker et al., 1997; Sanes and Donoghue, 1993). The oscillations terminate at movement onset and may well represent a top-down modulatory input from higher sensory areas (e.g., Lebedev and Wise, 2000). Furthermore, there is a large body of knowledge about delay-related spiking activity in motor cortical areas and its functional implication in sensorimotor integration and movement preparation (for a review, see Riehle, 2005). Finally, transient spike synchrony observed among individual neurons is remarkably well related to timing-related aspects of the behavioral task (Riehle et al., 1997; Kilavik et al., 2009) but does not depend on the mean firing rate of the participating neurons (Grammont and Riehle, 2003). However, only a few studies relate LFP oscillations to correlations of the spiking activity (Murthy and Fetz, 1996b; Nir et al., 2007). Reports in various brain areas demonstrate single neurons which selectively participate in oscillatory periods of the LFP by phase locking (Fries et al., 2001; Eckhorn and Obermueller, 1993; Baker et al., 1997; Destexhe et al., 1999), where occasionally the autocorrelations of the spike trains become oscillatory (Murthy and Fetz, 1996b; Lebedev and Wise, 2000). In conclusion, the apparent complexity of the simultaneous coding of neuronal activity for different aspects of motor cortical processing challenges the idea that LFP oscillations and the emergence of transient UEs are two reflections of only one single functional process performing the planning and preparation of movements.

We interpret the observed excess synchrony as a result of the specific activation of the observed neurons. An alternate hypothesis indicates that strong non-stationarities of the firing rates are the cause for false-positive detections of UE periods, which could explain the observed phase locking of UE if rates were co-modulated with the LFP oscillations cycles. To investigate this possibility, we reanalyzed the data by replacing the parametric distribution implementing the null hypothesis in the original UE analysis by a distribution derived by surrogates. The employed surrogate method (spike train dithering, see Grün, 2009) closely preserves the rate profiles and the inter-spike interval distributions, and leads to a conservative (Louis et al., 2010) classification of excess synchronous events. Despite the decreased sensitivity of the surrogate based method to detect excess synchrony, our analysis confirms the phase distributions for ISO, CC, and UE that are the essential finding of our study. Thus, they are not explained as a consequence of rate co-variations, but express excess synchrony as a reflection of coordinated network activity.

It is reasonable to assume that synchrony on a spike-by-spike level, and population oscillations expressed by the LFP both originate from network processes that involve the pulsed, synchronous co-activation of specific subsets of neurons. One may argue that in this case we should observe an even more distinct relationship between the two measures. However, our techniques to detect synchrony related to the activation of neuronal assemblies are limited. The Unitary Event analysis assesses indirectly which coincidences are more likely to originate from such activations based on the comparison of the time-resolved rate of observed and expected coincidences. Nevertheless, the set of UEs may be composed of coincidences resulting from assembly activation and a considerable fraction of chance coincidences (see estimate below). Therefore, although the difference in locking precision between significant (UE) and non-significant (CC) time segments seems small at first glance, in this light it is even more surprising that we are able to observe an enhanced phase locking for the UEs. The argument implies that the subset of coincidences caused by assembly activation has a tight locking to the LFP. This conclusion is supported by previous work demonstrating that coherent membrane potential oscillations do not generate synchronized output spikes, and that brief, simultaneous synaptic inputs to a cell are the likely drive for action potential generation (Poulet and Petersen, 2008).

Unitary Events prefer a particular phase of the LFP oscillation, a signal which is rather homogeneous across the motor cortex (Murthy and Fetz, 1996a; Rubino et al., 2006). This finding renders unlikely a model of processing where assemblies can be simultaneously active and still distinguished (multiplexed) by locking to different phases of the oscillatory cycle (e.g., Wommelsdorf et al., 2007). Moreover, in such a model the waxing and waning of the LFP oscillation would likely show phase shifts as different assemblies become active. Our results insinuate that neurons participate in different assemblies at different times (see also Riehle et al., 1997), but predominantly at the same phase of the LFP (cf., Singer, 1999). We observe the phenomenon in 20-30% of the neurons in agreement with estimates from other studies (e.g., Murthy and Fetz, 1996b). However, even in this category of neurons we can attribute only a fraction of spikes to assembly activation. One hypothesis is that the motor cortex is involved in parallel coding schemes, where synchronous assembly activity can be dissociated from the rate-based continuous-time coding.

To better understand the implications for the organization of cortical processing we consider a conceptual model where spikes of a neuronal assembly are locked to the LFP (Fig. 8) based on (i) the assumption that UEs reflect assembly activity (Riehle et al., 1997) and (ii) our observation that UEs have the strongest locking to the LFP. A potential mechanism is that assembly spikes originate from synchronous synaptic input to local groups of neurons. The simplest explanation for the finding that ISO and CC also exhibit locking, albeit weaker than UE, is that the spikes of a neuron are composed of a mixture of non-assembly (unlocked) and assembly spikes (locked). The latter are not identified as UE due to the lack of corresponding partner neurons in the recording (Fig. 8*A*). Consequently, the phase histogram of the ISO spikes is a superposition of the histograms of non-assembly and assembly spikes, with a factor $\gamma$ determining their ratio (Fig. 8*B*, top row). Chance coincidences are composed of spikes from independent sources (Fig. 8*B*, middle row) but the combinatorics of non-assembly and assembly spikes enhances the locking. Finally, periods identified as UE contain excess coincidences (Fig. 8*B*, bottom row) resulting from the activation of an assembly in which both neurons participate. Their relative contribution $\beta$ leads to an enhanced locking of UE compared to CC. The structure of the model allows us to derive minimal estimates of the parameters $\gamma$ and $\beta$ from the experimental phase histograms. We find that outside of UE periods $\gamma=13\%$ of the spikes of a neuron participate in an assembly, and $\beta=24\%$ of the coincidences in UE periods result from the joint participation in an assembly. Even though this is clearly a highly simplified model, it provides a first quantitative bridge between functionally relevant spike synchrony (Riehle et al., 1997; Singer, 1999; Maldonado et al., 2008) and the LFP as a robust mesoscopic measure of brain activity (Mehring et al., 2003).

Our results show that neuronal mass signals like the LFP convey specific information about network processes. We directly demonstrate in the brain of a behaving animal that the LFP is related to excess spike synchronization. Nevertheless, there is a substantial fraction of spikes without an apparent relationship to the LFP. Thus the two measures are observables of the same neuronal network but do not necessarily carry the same information. Taken together, we interpret our results as evidence that LFP (beta) oscillations, especially at high amplitudes, are reflections of the activation of neuronal assemblies which propagate a synchronous volley through the network. Complementing recent advances in tackling the experimental (Euston et al., 2007; Fujisawa et al., 2008; Nicolelis et al., 1997) and theoretical (Brown et al., 2004; Grün, 2009) difficulties in finding signatures of coordinated activity in spike data alone, these findings indicate how the LFP may provide a valuable additional source of information to characterize the neuronal population dynamics. With massively parallel recordings becoming available we may be able to disambiguate the superposition of multiple neuronal assemblies. This gives us confidence that by improving our understanding of the various components of the LFP signal we will eventually be able to use the LFP as an antenna delivering news from several communicating network stations.

# Funding

This work was supported by the Stifterverband für die Deutsche Wissenschaft; the Bundesministerium für Bildung und Forschung, Germany (BMBF, grant 01GQ0413 to BCCN Berlin); the Helmholtz Alliance on Systems Biology; the French National Research Agency (ANR-05-NEUR-045-01); the Deutscher Akademischer Austauschdienst (DAAD); The Research Council of Norway (eScience Programme) and the European Union (grant 15879, FACETS).

# Acknowledgments

We thank Moshe Abeles, Walter Freeman, and George Gerstein for valuable comments on an earlier version of the manuscript

# Figure Legends

**Figure 1. Characteristics of LFP and spiking dynamics.** (*A*) Two single-trial LFPs recorded simultaneously (gray) at different electrodes (during long trials with movement to the right in the SELF task). Superimposed are the beta-filtered (10-22 Hz) signals (red) and their instantaneous oscillation phase (black lines). The histogram visualizes the phase differences between the two signals across all time bins. (*B*) Spike raster of one example neuron recorded in parallel to the LFP shown above. (*C* and *D*) Neither the trial-averaged inter-spike interval distribution (*C*) nor the normalized auto-correlograms (*D*) indicate an oscillatory nature of the neuron. (*E*) The cross-correlogram with a different neuron recorded in parallel (neuron 1 in supplemental Fig. S1) remains flat. Red lines indicate mean (solid) and 5% confidence intervals (dashed) of cross-correlograms obtained from surrogate spike trains where each spike was jittered uniformly in window of ±20 ms around its original position.

**Figure 2. Sketch of the analysis.** Spikes of two neurons (yellow background) and an LFP are recorded from electrodes separated by approximately 400 µm (right). Spikes are classified as isolated (ISO, gray), chance coincidence (CC, cyan), or Unitary Event (UE, red) depending on their precise synchronization with a spike of a second neuron recorded in parallel. In contrast to CCs, UEs identify coincidences in transient epochs where the high number of observed coincidences (top left) significantly exceeds the prediction based on the firing rates (in practice, coincidences are counted across trials, which is omitted here for illustrative purpose). In UE epochs, synchrony

between both neurons in excess of the chance contribution is explained by their specific co-activation in a neuronal ensemble, termed assembly. Two assemblies are sketched in green and blue but the recorded neurons participate only in the green one. We investigate the relationship of the two types of observed spike synchrony (CC and UE) to the LFP population signal as a monitor of brain processing.

**Figure 3. The magnitude of the spike-triggered average (STA) depends on the occurrence of synchronized spiking activity.** (*A*) STA of the LFP averaged over all 123 neurons (n=297484 spikes total) for the three disjunct sets of spikes. The left panel compares STAs of ISO (dark gray curve, n=240455) to CC (cyan curve, n=44867). To account for the difference in variability due to sample sizes, the STA of ISO is repeatedly recomputed using only 44867 random trigger spikes. The light gray band encloses at each point in time 95% of all recomputed STAs. The middle and right panel compare STAs of UE (red curve, n=12162) to ISO and CC, respectively. (*B*) Relative number of neurons per animal (vertical) with the STA of one spike set exceeding (in area) the STA of the other set (horizontal, color codes). The STA of the first set qualifies as larger if it exceeds the other STA in 50% of 1000 recomputations (superimposed darker bars: 95%, i.e. $\alpha$=5%). (*C*) The four bars distinguish STAs obtained for neurons with the same number $N_p$ of partner neurons used in coincidence detection. Same criteria (50%, both animals) as in *B*. (*D*) The correlation of LFP amplitude and spike rate is not significant ($\alpha$=0.01, coefficient $R$).

**Figure 4. LFP-spike phase coupling reveals locking increase for coincidences.** (*A*) Determination of phase and amplitude (example neuron). Top: single LFP trial;

middle: trial-averaged power spectrogram. The beta activity during the preparatory period (PP, between PS and AT) disappears with movement (Mvt). Bottom: Phase (green) and amplitude (blue) of the beta-filtered LFP (upper trial shown in the top graph) extracted at the spike times (ticks). Resulting spike-triggered phase distributions (green) are characterized by their circular standard deviation $\sigma$. Same neuron as in Fig. 1. (*B*) Percentage of neurons in ISO (gray curve), CC (cyan), and UE (red) with a circular standard deviation of the phase distribution below $\sigma$ (horizontal axis). For the average $\sigma_T$=1.97 of the set of significantly locked neurons (all spikes, $\alpha$=0.05) the percentages are also shown as bars. (*C*) Comparisons of the circular standard deviations $\sigma$ of the three sets in the individual neurons: ISO vs. CC (top, n=291), ISO vs. UE (middle, n=142), and CC vs. UE (bottom, n=136). Each dot represents one neuron in one experimental configuration. The percentages show the relative number of data points above the diagonal. The light (dark) gray ellipse covers 2 (1) standard deviations of the sample variance (outlined ellipse: surrogate data ISO vs. CC with shuffled ISIs).

**Figure 5. Phase and amplitude distributions in a single neuron.** Same neuron as in Figs. 1 and 4. All distributions are normalized to unity area and are shown separately for ISO (left), CC (middle), and UE (right). (*A*) The modulation of the phase distribution increases from left to right. Phase $\pi$ is the location of the trough of the LFP oscillation. The black curve in the middle and the right panel is the expected phase distribution of coincidences predicted from the phase distributions of the contributing neurons (see Methods). (*B*) Simultaneously to the increased locking, the amplitude distribution shifts to higher values.

**Figure 6. Relation of spike synchrony to the interplay of phase and amplitude.**
(*A*) Joint histograms of the phase and amplitude for ISO (left), CC (middle), and UE (right) pooled across the population (color bars indicate counts; phase π indicates LFP troughs). The top and left projections display the phase and amplitude distributions, respectively. The top middle and top right graph compare the phase distribution to the distribution shown in the graph to the left: The shaded areas enclose at each phase 95% of 1000 phase distributions randomly chosen from the set to the left with the same number of spikes as in the current set. Black curves are the predictions based on the phase distributions of the individual neurons. The histograms include the neurons that have a minimal spike count (total of 25 spikes and a mean rate of 5 Hz per trial) and for which the phase distribution of all spikes is significantly locked (α=0.05). (*B*) Phase distributions of the three sets, considering only 50% of spikes at the highest LFP amplitudes (above dashed black line in *A*).

**Figure 7. Influence of oscillation magnitude on locking of spikes to LFP.** (*A*) Spikes in periods with an LFP magnitude (i.e. envelope of LFP, light gray curve) above a certain threshold (dashed line) are termed the 'Hi' set (light gray ticks) and the remainder the 'Lo' set (dark gray ticks). (*B*) Separation of spikes into Hi and Lo for the same example neuron as in Figs. 1, 4, and 5. Spikes are rank ordered according to LFP magnitude; the histogram on the right shows the distribution of the respective magnitudes. The threshold $\theta$ is defined as the relative number of spikes labeled as Lo. The dark gray arrow illustrates a threshold choice of $\theta$=0.5, and corresponds to a data dependent relative amplitude (light gray arrow). Spikes at extremely low LFP amplitudes (lowest 10%) do not enter the analysis. (*C*) Percentage of neurons with significant (Rayleigh test, α=0.05) phase-locking of the Hi spikes (light gray curve) and of the Lo spikes (dark gray curve) as a function of magnitude threshold. Even for large $\theta$ (0.8) the set of Hi spikes shows significant locking in 36% of the neurons, although it consists of only few spikes. The dashed line shows as a reference the percentage (39%) of locked neurons computed if spikes are not separated into Hi and Lo (i.e. all spikes). Thus the locking of neurons is mainly explained by the locked Hi spikes, and their locking is approximately independent of $\theta$.

**Figure 8. A conceptual model relating increased LFP locking and assemblies.** (*A*) Sketch of the LFP (top) and the simultaneous spiking activity of five neurons (middle), of which only two are recorded (yellow background). Based on the latter, time periods where coincidences occur at chance level (non-UE, left) are distinguished from those with excess synchrony (UE, right). Each spike is either part of an assembly of co-active neurons (green) or not (black). In this simplified scenario, one assembly is active on the left, and a different one on the right; both observed neurons contribute to the latter. Only assembly spikes exhibit locking to the LFP, expressed by a non-uniform phase distribution $p(\phi)$ (green). (*B*) Two ratios $\beta$ and $\gamma$ determine the composition of the phase distributions for ISO, CC, and UE (left) of assembly and non-assembly spikes. $\gamma$ determines the overall probability that a spike is part of an assembly activation (top, ISO). $p_{CC}(\phi)$ (middle) results from the combinatorics of two independent spike trains (ISO). $p_{UE}(\phi)$ (bottom) differs from $p_{CC}(\phi)$ by the relative excess $\beta$ of assembly spikes in UE periods. A conservative (minimal) estimate of $\beta$, i.e., maximally locked $p^2(\phi)$, is obtained by substituting $p_{UE}(\phi)$ and $p_{CC}(\phi)$ in the bottom equation by the experimental distributions. $\gamma$ is determined from either of the top two equations by using $p(\phi)$.

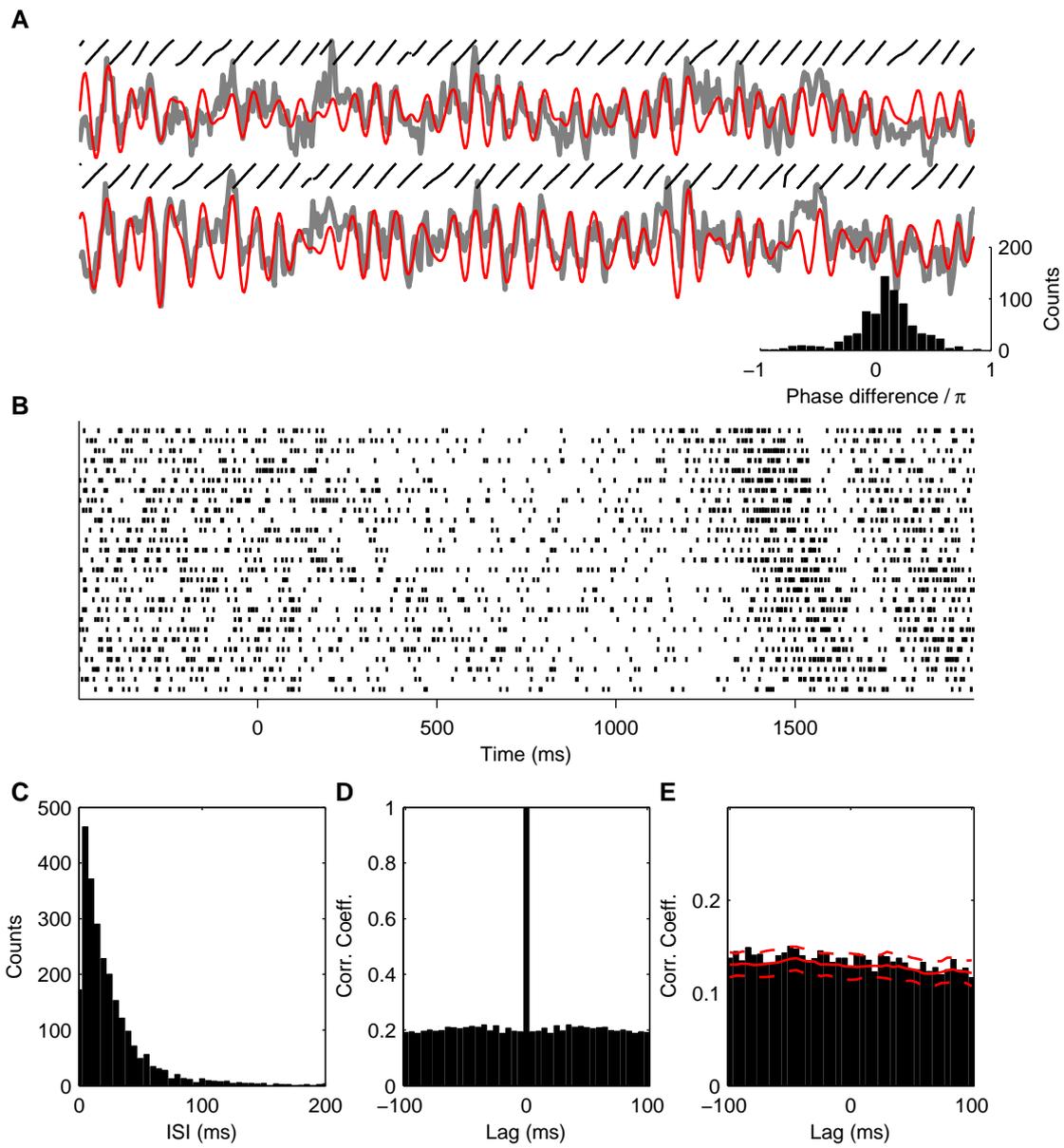

Figure 1

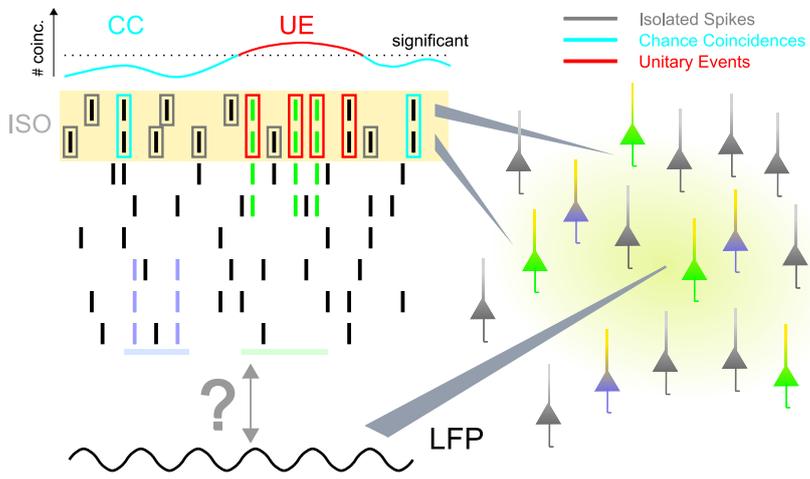

Figure 2

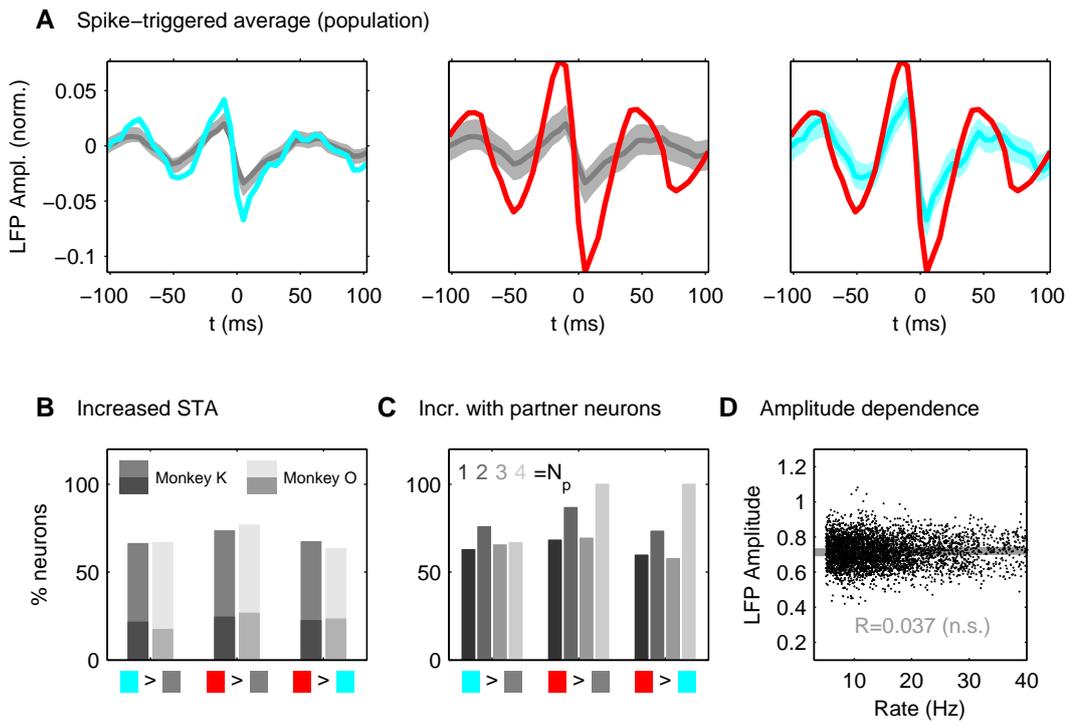

Figure 3

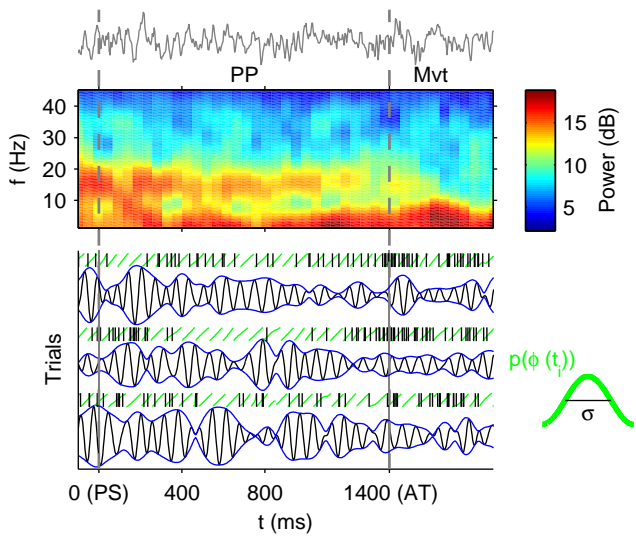
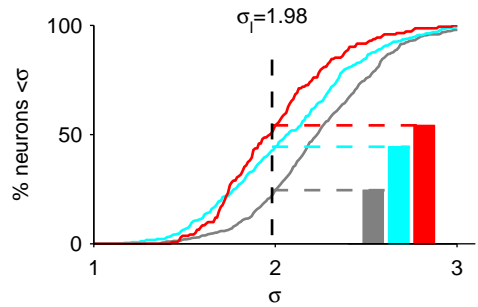
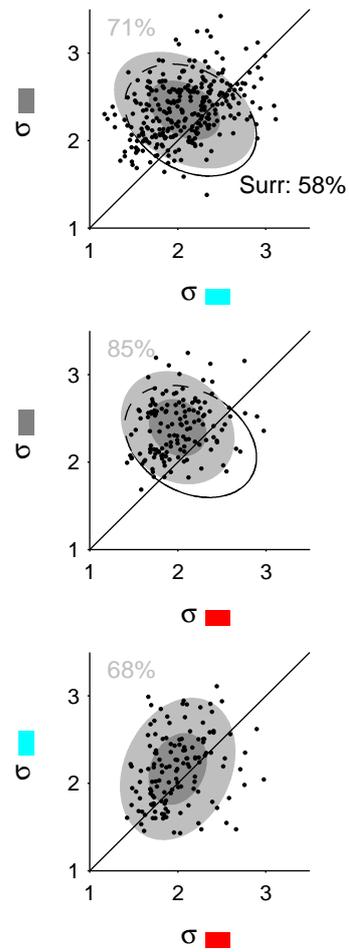

Figure 4

**A**  Phase distribution (example neuron)

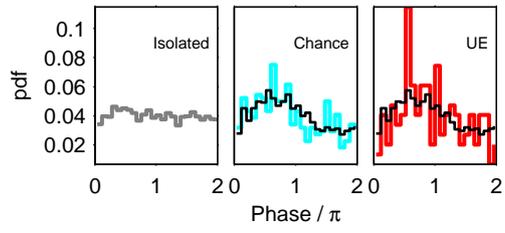

**B**  Amplitude distribution (example neuron)

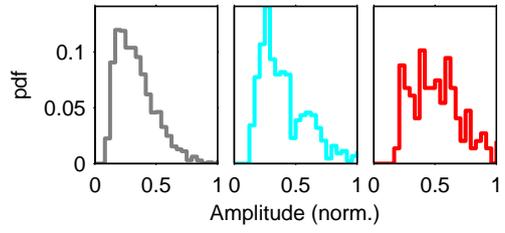

Figure 5

**A** Phase and amplitude distributions

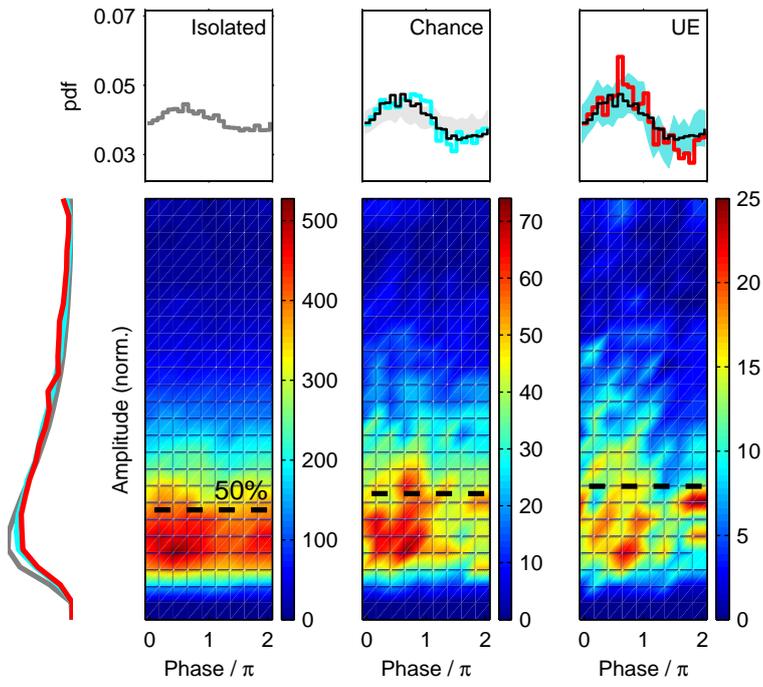

**B** Spikes at high LFP amplitudes

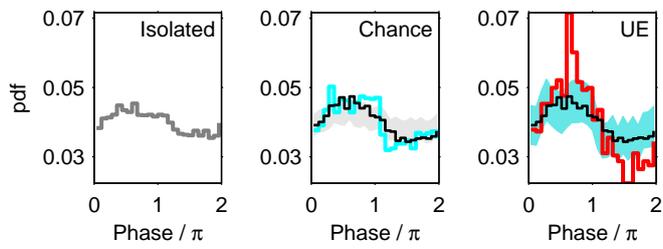

Figure 6

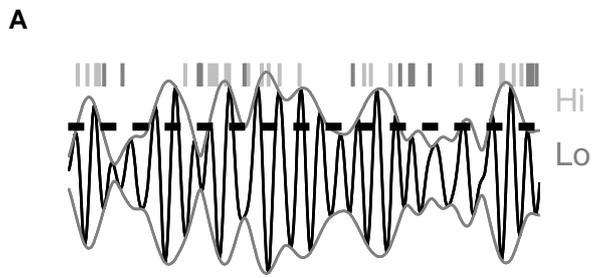

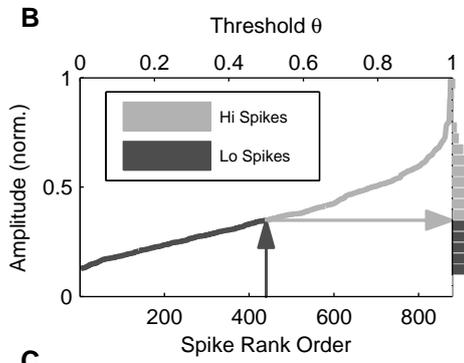

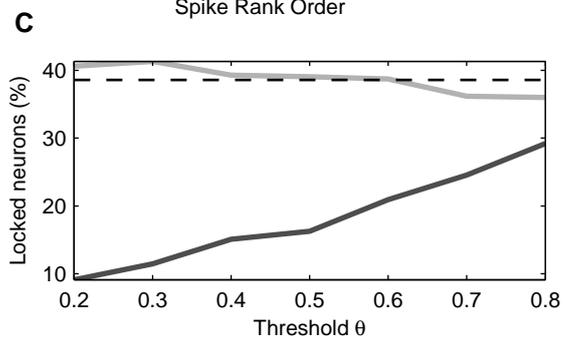

Figure 7

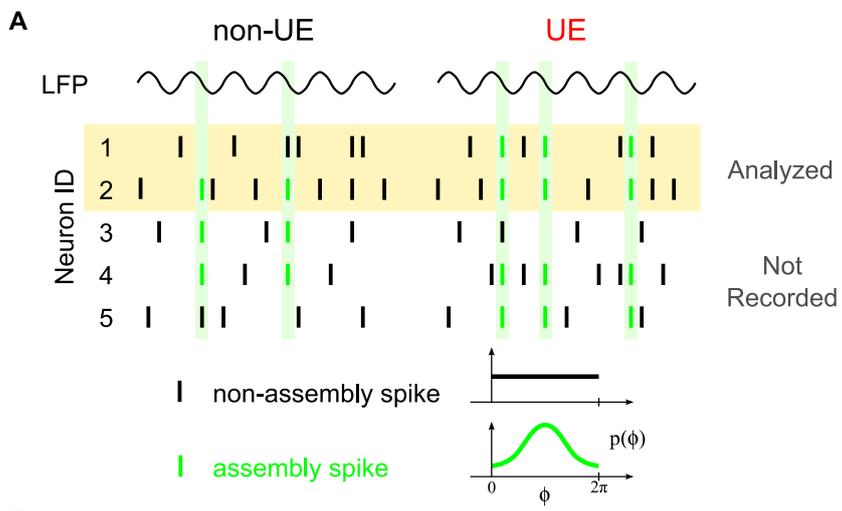

Figure 8

# The Local Field Potential Reflects Surplus Spike Synchrony

# Supplemental Information

## Supplemental Figure Legends

**Supplemental Figure S1. Detection of Unitary Events.** (*A*) Spike rasters for the same neuron (neuron 6) shown in Figs. 1, 4, and 5 and one simultaneously recorded neuron (neuron 1). Each line in the rasters corresponds to one trial. Simultaneously recorded activities of the two neurons are shown on lines of the same height in the respective raster. Spikes are indicated by black dots, coincident spikes and Unitary Events are surrounded by a cyan or red square, respectively. Data shown are recorded during the self-paced task with long time delay (see Methods for experimental details). The corresponding behavioral events are marked in the rasters with differently colored filled circles: occurrence of the preparatory stimulus PS (dark red), allowed movement time AT (light blue), movement initiation (dark blue) and end of movement (dark green). (*B*) Firing and coincidence rates. The firing rates of the two neurons are shown in dark gray (neuron 6) and light gray (neuron 1), together with the rate of the empirical coincidences (light cyan) and the coincidence rate expected from the neurons' firing rates (dark cyan), calculated as the sum of the trial-by-trial rates. All rates are estimated in sliding windows of 100 ms width shifted by 0.1 ms. (*C*) Significance of empirical coincidences. The joint surprise (dark gray curve) results from the comparison of the empirical and the expected coincidence counts.

Significant excess coincidences (i.e. UEs) are detected if the joint surprise is larger than the 5% level (dashed line). For comparison, the 1% level is also indicated (dotted line). UEs are found during a short period before PS occurrence, shortly after PS, and at 600 ms after PS. The latter is one of the short delay times that monkey was exposed to in parallel to the shown delay scheme. Note that although there is a considerable increase of coincident events in relation to the arm movement, they occur at chance level.

**Supplemental Figure S2. Relationship of LFP and synchronized spiking behavior in a single neuron and LFP-triggered PSTHs of synchronized activity.** (*A*) STA of the LFP (filtered between 2-80 Hz to remove DC components) of one neuron (same neuron as in Figs. 1, 4, 5, and S1) for three disjunct sets of trigger spikes: not coincident with spikes from simultaneously recorded other neurons (isolated spikes, ISO, gray), involved in coincidences (within 3 ms) predicted by rate (chance coincidences, CC, cyan), and involved in significant coincidences (Unitary Events, UE, red). The left panel compares the STA of ISO (dark gray curve, n=4098) to the STA of CC (cyan curve, n=506). To account for the difference in variability due to sample sizes, the STA of ISO is recomputed using only 506 random trigger spikes. The light gray band results from the superposition of 1000 re-computations of which 95% are enclosed by the dashed curves at each point in time. Similarly, the middle and right panel compare the STA of UE (red curve, n=177) to the STA of ISO and CC, respectively. (*B*) Bottom: Population-averaged LFP-triggered histogram of ISO (left), CC (middle), and UE (right). The trigger times are the largest local maxima of the LFP that are separated by a minimum distance of 33 ms. The spikes of a neuron are

triggered on exactly one LFP channel. Top: LFP averages for each neuron contributing to the histogram (light gray curves) based on the same trigger. The dark gray curve is the average of the single neuron LFP averages.

**Supplemental Figure S3. The increased locking of UEs is independent of the overall degree of locking of the neuron.** (*A*) Fraction of neurons exhibiting (threshold of 25 spikes) ISO, CC and UE separately for the sets of strongly (left) locked and weakly (right) locked neurons (criterion: surrogate test ($\alpha=0.05$) on original spike train containing all spikes). (*B*) Percentage of neurons with a locking stronger than $\sigma_l$ in each of the two groups (strongly and weakly locked). For the selected value of $\sigma_l=1.98$ (average locking strength of strongly locked neurons) the percentages are shown as bars.

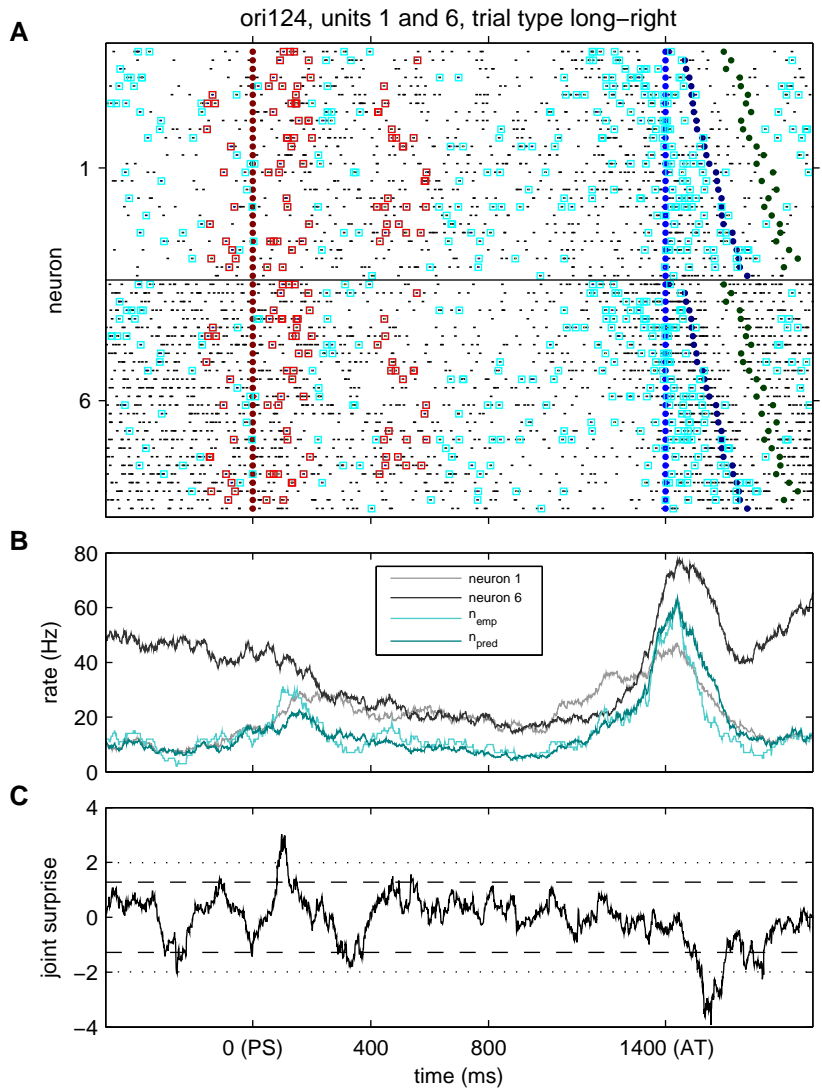

Supplemental Figure S1

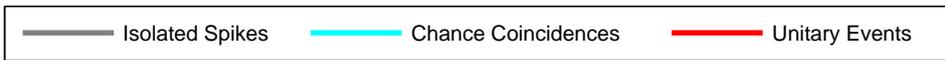
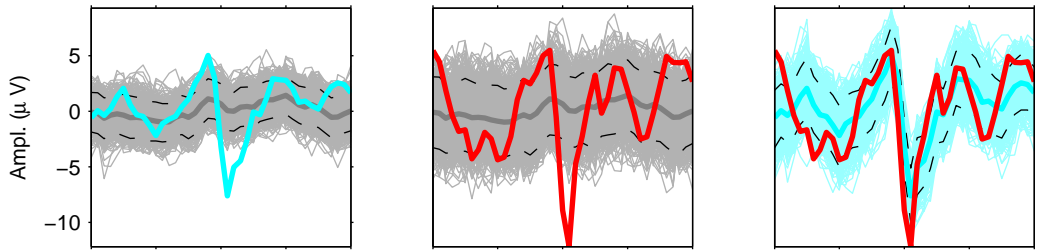

**A** Spike−triggered average (example neuron)

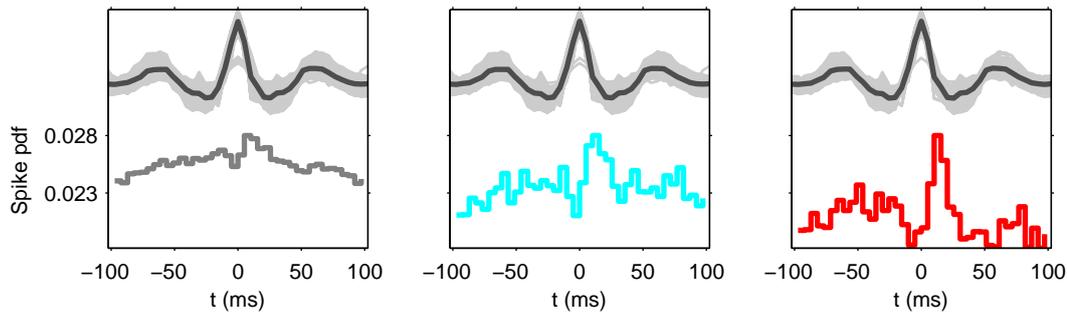

**B** LFP−triggered spike histogram

Supplemental Figure S2

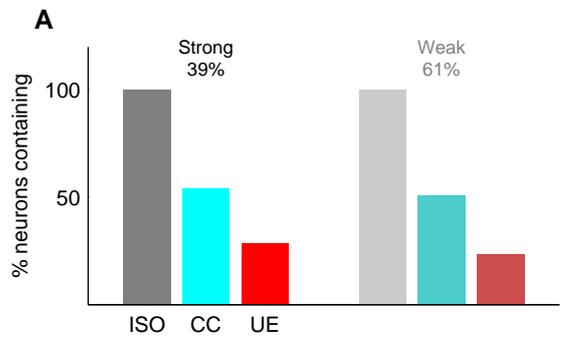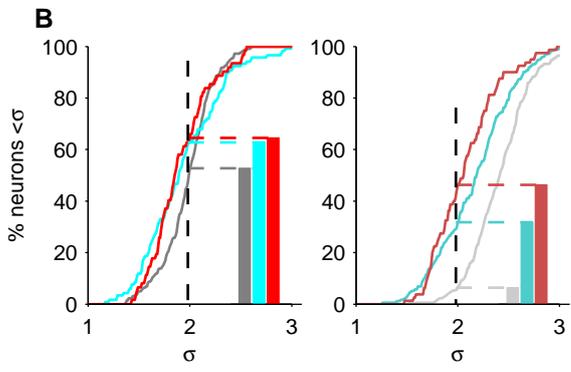

Supplemental Figure S3